\documentclass[11pt,a4paper]{article}
\pdfoutput=1
\usepackage{jheppub}
\usepackage{subfigure}
\usepackage{slashed}

\def\TeV{\textrm{Te\kern -0.1em V}}
\def\GeV{\textrm{Ge\kern -0.1em V}}
\def\GeVcc{\textrm{Ge\kern -0.1em V}/$c^2$}
\def\tautau{\ensuremath{\tau^+ \tau^-}}
\def\ra{\ensuremath{\rightarrow}}
\def\z0{\ensuremath{Z^0}}
\def\pt{\ensuremath{p_{\mathrm{T}}}}
\def\MET{\ensuremath{\slashed{E}_{\mathrm{T}}}}
 
\def\pimode2{\ensuremath{h^{\pm}\,\nu}} 
\def\pimode{\ensuremath{h^{\pm}\,\nu_\tau}} 
\def\pionmode{\ensuremath{\pi^{\pm}\,\nu_\tau}} 
\def\rhomode2{\ensuremath{h^{\pm}\,\nu~n h^{0}}} 
\def\rhomode{\ensuremath{h^{\pm}\,\nu_\tau~n h^{0}}} 
\def\lmode{\ensuremath{\ell^{\pm}\,\nu_\tau\,\overline{\nu}_{\ell}}}
\def\Evis{\ensuremath{E^\mathrm{vis}}}
\def\mvis{\ensuremath{\mathrm{M}_\mathrm{vis}}}
\def\meff{\ensuremath{\mathrm{M}_\mathrm{eff}}}
\def\RF{\ensuremath{\mathrm{RF}}}
\def\logL{\ensuremath{\log\mathcal{L}}}
\def\eboost{\ensuremath{2E^{1}_\RF}}
\def\mboost{\ensuremath{M_\mathrm{boost}}}
\def\bt{\ensuremath{\beta_\mathrm{T}}}
\def\bz{\ensuremath{\beta_\mathrm{z}}}
\def\bzt{\ensuremath{\beta_\mathrm{z,T}}}
\def\dphi{\ensuremath{\Delta\phi^{\mathrm{jets}}}}
\def\xyz{XYZ}
\newcommand{\tauola}{\texttt{TAUOLA}}
\newcommand{\pythia}{\texttt{Pythia8}}
\newcommand{\avg}[1]{\left< #1 \right>}

\title{A Method to Estimate the Boson Mass and to Optimise Sensitivity to 
Helicity Correlations of \tautau\ Final States}

\author{Peter L. Rosendahl,}
\emailAdd{Peter.Rosendahl@ift.uib.no}
\author{Thomas Burgess}
\emailAdd{Thomas.Burgess@ift.uib.no}
\author{and Bjarne Stugu}
\emailAdd{Bjarne.Stugu@ift.uib.no}

\affiliation{University of Bergen, Department of Physics and Technology, N-5020 
Bergen, Norway}

\abstract{
In proton-proton collisions at LHC energies, \z0 and low mass Higgs bosons 
would be produced with high and predominantly longitudinal boost with respect 
to the beam axis. This note describes a new analysis tool devised to handle 
this situation in cases when such bosons decay to a pair of $\tau$-leptons. The
tool reconstructs the rest frame of the \tautau\ pair by finding the boost that 
minimises the acollinearity between the visible $\tau$ decay products. In most 
cases this gives a reasonable approximation to the rest frame of the decaying 
boson. It is shown how the reconstructed rest frame allows for a new method 
of mass estimation. Also a considerable gain in sensitivity to 
helicity correlations is obtained by analysing the $\tau$-jets in 
the reconstructed frame instead of using the laboratory momenta and energies, 
particularly when both $\tau$-leptons decay hadronically.
}

\notoc
\begin{document}
\maketitle
\flushbottom

\section{Introduction}
Experimental precision data strongly suggest that the Higgs boson should be 
relatively light. While experimental data exclude Standard Model Higgs, $H$, 
masses below 114.4 \GeVcc, the lightest neutral supersymmetric 
Higgs can have a mass as low as 92.8~\GeVcc~\cite{PDG}, which is very close to 
the mass of the \z0. In the mass range below 160~\GeVcc\ the \tautau\ decay 
mode is of particular importance in searches for neutral $H$, as demonstrated 
in Tevatron searches~\cite{D0CDF}. In this mass region, $H$ production 
in proton-proton (pp) collisions at LHC energies is dominated by 
gluon-gluon fusion~\cite{ATL-2010-015}, and a search for $gg \rightarrow H 
\rightarrow \tautau$ production at the LHC is very important, as emphasised 
in~\cite{DJOUADI}. 

Together these facts strongly motivates a careful study of the \tautau\ system 
as produced at the LHC, and the development of analysis tools that enhance 
sensitivity to measure properties of $H$ bosons. 
Furthermore, since \z0\ production is an irreducible background for this 
channel it is also important to fully understand the behaviour of 
\z0\ra\tautau\ events and to look for possible contributions from other 
processes in a selected sample of \tautau\ pairs. The sensitivity of such a 
study can be enhanced by exploiting the fact that the parity violating 
$\tau$ decays carry helicity information. Thus, $\tau$-leptons can be used as 
spin analysers, and provide information that could be helpful in distinguishing 
the existence of a scalar $H$ in a background of \z0 decays.

In~\cite{WAS} some variables that could be used for a spin analysis are
proposed. However, these are all defined in the rest frame of the decaying
boson, a frame that cannot be found in any straight forward manner because of
the escaping neutrinos produced in the $\tau$-decays. 

Methods have been proposed to assign the measured missing transverse
energy, $\MET$, to the neutrinos to obtain unbiased estimates of
\tautau\ mass~\cite{RKELLIS,MMC}. The resulting estimates of the 
neutrino momenta also gives access to an approximation of the rest
frame of the decaying boson, but implications of this beyond mass
estimation is not discussed. Further, these methods only work in a fraction 
of the events with a suitable topology whereas the technique proposed
in this paper is applicable for all event topologies. 

This note presents a new simple algorithm that finds an approximation to the 
rest frame of the decaying boson. This will be shown to result in an improved 
sensitivity to spin, and to an alternative variable that could be of use for 
the estimation of the mass of the decaying boson.

\section{Reconstructing the \texorpdfstring{\tautau}{tau+tau-} rest frame}
\label{sec:method}
Because of their short life time $\tau$-leptons are typically studied 
indirectly through their decay products. Due to the one or two neutrinos 
produced in the decay, the 4-momentum of the visible $\tau$-decay products, the 
$\tau$-jet, has a broad spectrum, often carrying less that half the energy 
of their mother $\tau$. Therefore, reconstructing the rest frame, \RF, of a 
particle decaying into a \tautau\ system is in principle not possible.
 
However, for a sufficiently massive particle decaying into a \tautau\ pair, the 
directions of the $\tau$-jets are expected to be close to that of their 
mother $\tau$-leptons. Thus, one expects the $\tau$-jet pair to have just only 
a slight deviation from being back-to-back in the rest frame of the massive
particle. It is useful to define the 
\emph{acollinearity} -- the angular deviation of the $\tau$-jets from being 
back-to-back. This acollinearity is a well defined positive number in any 
frame of reference. Henceforth, $\alpha$ will denote the acollinearity between 
two $\tau$-jets, and furthermore $\alpha_\RF$ the acollinearity in the heavy 
particle rest frame.

Simulations show that at 7~\TeV\ pp-collisions, the \z0 and a light $H$, 
are mostly produced with a high and predominantly longitudinal boost,
\bz. Thus, measured 4-momenta will deviate significantly from those in 
the heavy particle \RF. Exploiting that $\alpha_\RF$ should be small 
and assuming that \bt\ is small, the method proposed in this paper 
consists in searching for the \bz\ that minimises $\alpha$. When 
restricting $\beta$ to be parallel to the beam axis, the $\alpha$ as a function 
of $\beta$ has a single minimum and hence can be reliably minimised. 

The method can also extended to look for the transverse component of the boost 
if one has a good estimate of the direction of the heavy particle in the 
transverse plan. The transverse direction of the heavy particle, can be 
estimated by summing up the transverse momenta of the $\tau$-jets and the 
reconstructed missing transverse energy, $\MET$. Our algorithm minimises 
$\alpha$ in two steps; first, by varying the longitudinal boost, \bz, 
secondly when a suitable minimum is found, the transverse boost is varied 
along the found transverse direction. In both steps the $\alpha$ is 
1-dimensional function with a single global minimum and no secondary minima. 

In this paper, we will denote the two search strategies as the Z- or 
\xyz-method depending on whether we only will try to estimate the longitudinal 
component of the boost of the heavy particle or estimate the transverse 
component as well. 

Since particle directions are well measure quantities these methods are robust 
and can be applied in every collision event for any pair of measured particles 
or jets. 

\subsection*{Performance}
The performance of the methods were studied using \tautau\ pairs from simulated 
\z0 and 114~\GeVcc\ gluon-gluon fusion $H$ events produced in 7~\TeV\ 
pp-collisions. Anticipating that backgrounds could be difficult to
handle in multi-prong $\tau$ decays, only decays with exactly one
charged particle in the final state were considered here. 
However, the inclusion of $\tau$ decays with more than one charged
hadron is not expected to change the performance of this method  
significantly.  $10^7$ \z0 and $H$ events were generated with
\pythia~\cite{PYTHIA} and subsequent $\tau$-decays were done with 
\tauola~\cite{DAVIDSON} to correctly include spin and polarisation
effects.  Detector resolution effects were not included. However, an angular cut 
requiring the two jets to have $|\eta|<2.5$ was imposed to match the acceptance 
of a typical tracker in an LHC experiment.

\begin{figure}[htbp]
  \begin{center} 
    \subfigure[Polar angle, $\theta$, of the \z0 and 114~\GeVcc\ $H$ bosons. 
    Typically the bosons are produced at small angles close to the beam axis.]{
      \includegraphics[width=0.45\textwidth]{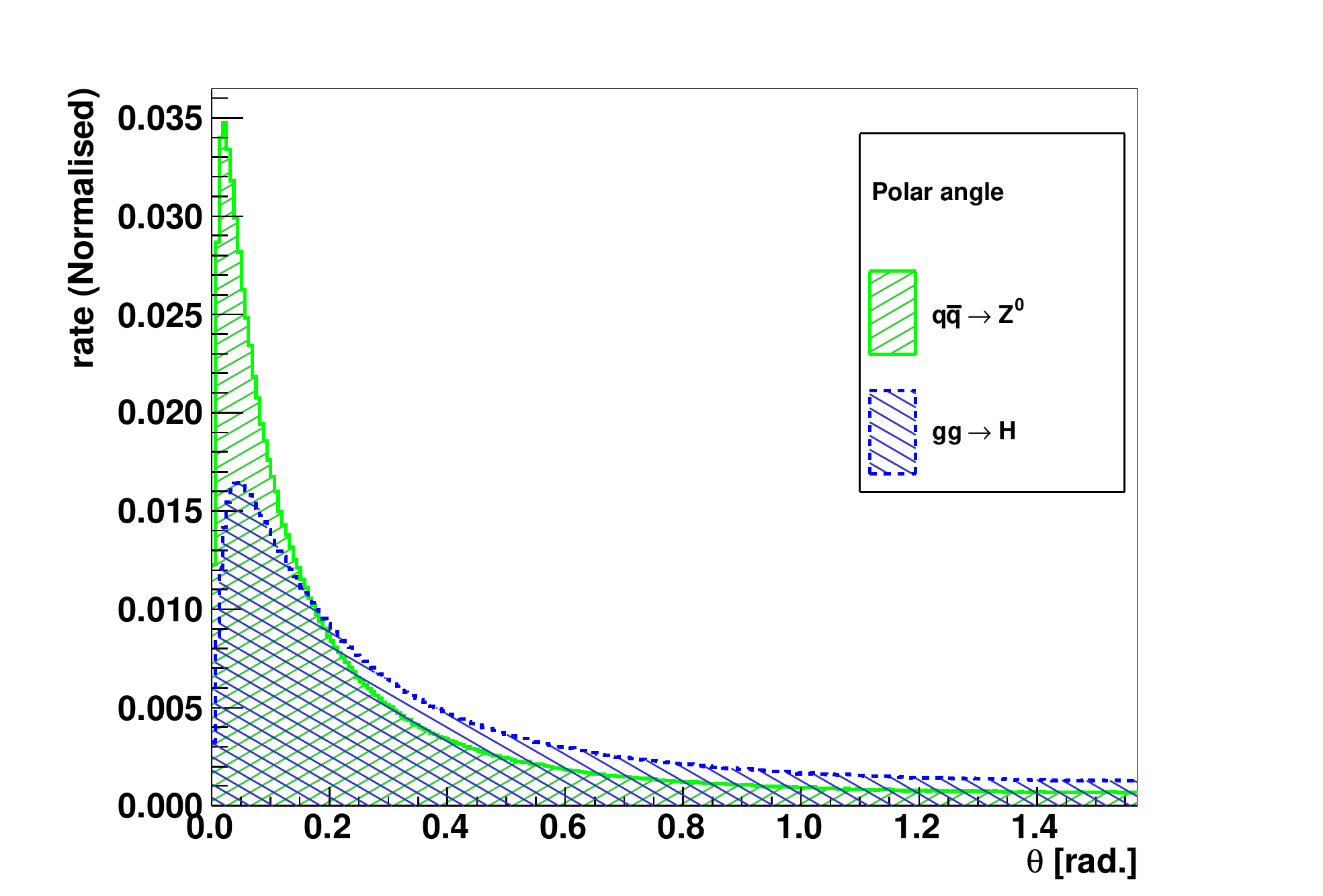}
      \label{fig:boost_theta}
    }
    \hspace{0.5 cm}
    \subfigure[Acollinearity, $\alpha$, for $\z0\ra\tautau$ in the generated 
    and reconstructed $\RF$s and in the 
    laboratory frame of the \z0.]{
      \includegraphics[width=0.45\textwidth]{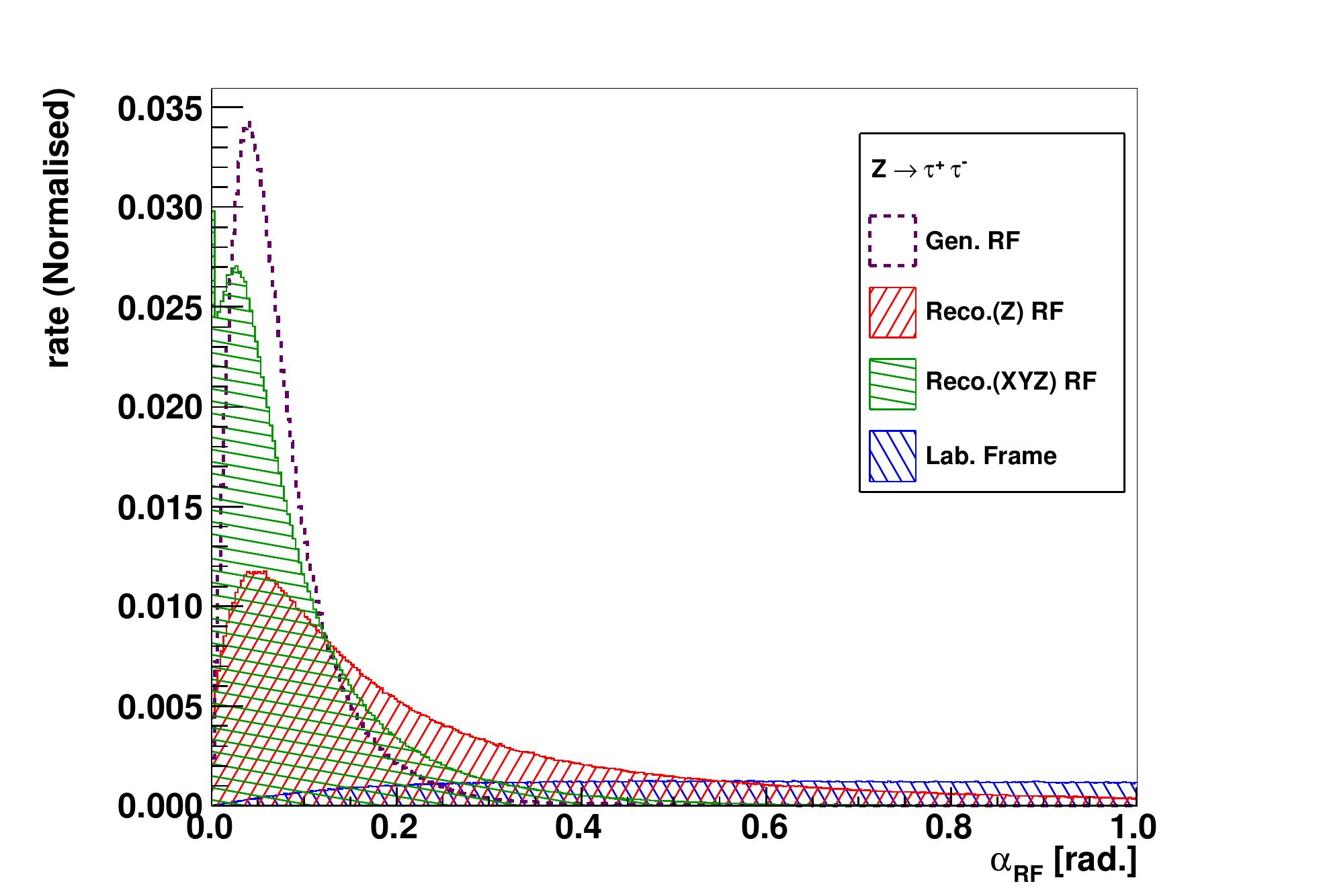}
      \label{fig:z0aco}
    }
    \caption{Angular distributions for simulated heavy boson decays.}
  \end{center}
\end{figure}
As shown in figure~\ref{fig:boost_theta} the polar angle distributions for both
$H$ and \z0\ decays are predominantly parallel to the beam axis. Therefore,
assuming a boost direction parallel to the beam seems to be a reasonable guess
for a very large fraction of the events and we therefore only expect the 
\xyz-method to be slightly better in reconstructing the rest frame than the 
Z-method for these events. In figure~\ref{fig:z0aco} it is seen that the true 
$\alpha_\RF$ is concentrated at small values, with $\avg{\alpha_\RF}=4.5^\circ$ 
for \z0 events, while the acollinearity distributions in the reconstructed 
frames are peaked towards low acollinearities as expected. The effect of working 
in the collinear frame ($\alpha=0$) instead of in the true \RF\ should have a 
small effect if the direction of the boost is correctly estimated. In
the collinear frame closest to the true \RF, the momenta of the two
$\tau$-jets will deviate with a factor $\cos(\alpha_\RF)$ from the
truth, meaning a typical deviation of $0.3\%$.

\begin{figure}[htbp]
  \begin{center}
    \includegraphics[width=0.65\textwidth]{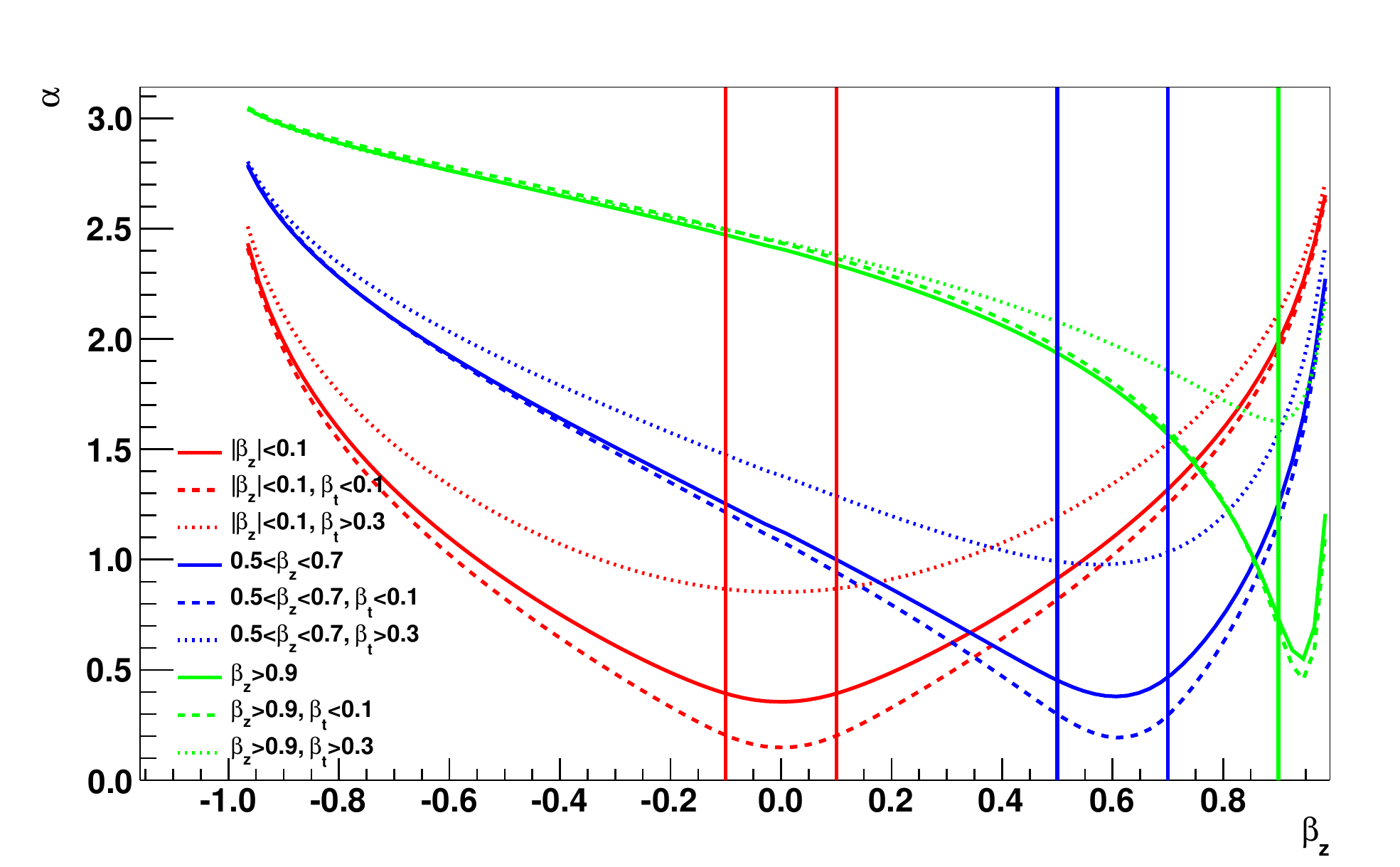}
    \caption{Average acollinearity as a function of attempted longitudinal boost 
      for various ranges of true boost.}
    \label{fig:betaVsAco}
  \end{center}
\end{figure}
Since the $\alpha$ is always positive, it is easy to minimise with respect to 
the applied $\bzt$ of the two tau jets. For a selection of ranges in 
the generated $\beta$ of a produced \z0-boson,
figure~\ref{fig:betaVsAco} shows how,  on average, $\alpha$ varies as
a function of an applied longitudinal boost to  the two jets. The
minimal acollinearity is always at a longitudinal boost that is close
to the generated one. The actual value of $\alpha$ is
reflecting to which extent there might be a transverse boost present,
ignored in figure~\ref{fig:betaVsAco}. 

\begin{figure}[htbp]
  \begin{center}
    \subfigure[Vectorial difference between the generated and reconstructed 
    $\RF$.]{
      \includegraphics[width=0.45\textwidth]{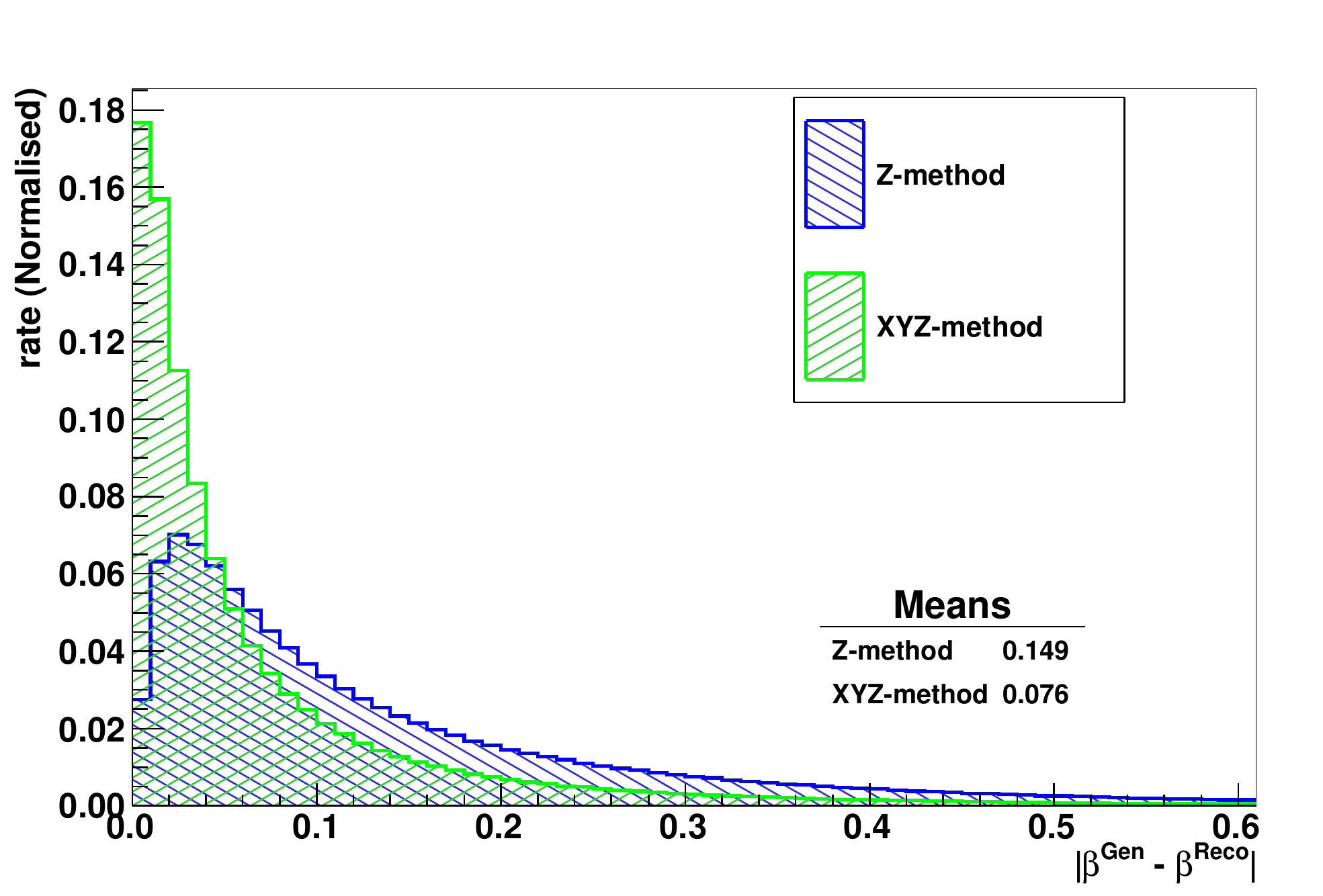}
      \label{fig:betaDiff}
    }
    \hspace{0.5 cm}
    \subfigure[Difference in magnitude between the generated and reconstructed 
    $\beta$.]{
      \includegraphics[width=0.45\textwidth]{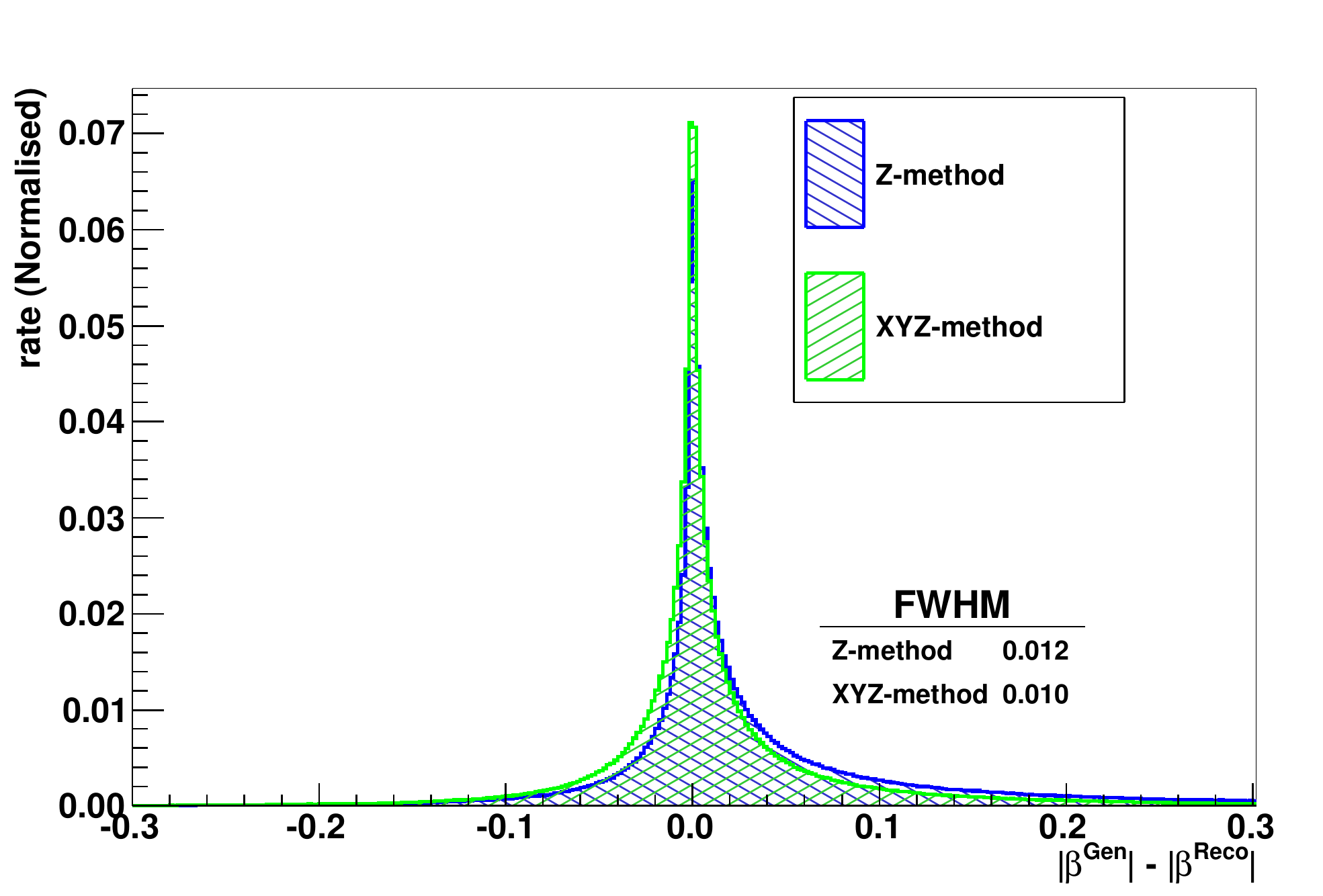}
      \label{fig:betaDiffMag}
    }
    \caption{Comparisons of both reconstructed and generated rest frames in 
      $\z0\to\tautau$ 
    events.}
    \label{fig:betarec}
  \end{center}
\end{figure}
To estimate the goodness of the reconstruction of the \RF, the vectorial 
difference $|\beta_\mathrm{gen}-\beta_\mathrm{reco}|$ is shown in 
figure~\ref{fig:betaDiff} 
for both reconstruction methods. The reconstructed $\RF$s are seen to be 
closer to the generated \RF\ for the \xyz-method as expected. 
Figure~\ref{fig:betaDiffMag} shows the difference in magnitude between the 
generated and reconstructed boost with a FWHM of the distribution of $0.010$ 
for the \xyz-method. 

In conclusion both methods reconstructs the boost very reasonably, with a 
distribution around the true value with a mean deviation around $0.076$ for 
the \xyz-method and $0.149$ for the Z-method, when averaging over all events. 
Due to the exclusion of the transverse component, a small offset away from 
zero in the most probable value is seen in the Z-method. 
Although figure~\ref{fig:boost_theta} shows that the bosons are mostly boosted 
longitudinally, improvements are found when also estimating the transverse 
components of the boost. 
However, these improvements will be very dependent on the ability to
estimate the transverse direction of the bosons correctly in the
detector. Therefore the two methods act complementary; for events
with a small transverse boost and low precision on the \MET\ the
Z-method should be applied whereas for events with good 
precision in \MET\ and a high transverse boost improvements will be 
gained by using the \xyz-method.

\section{Estimating the Boson Mass}
Due to the neutrinos in the $\tau$ decays, the visible invariant mass, \mvis, 
distribution peaks far below the real mass of the resonance produced. A method 
correcting for this is the so called 
\emph{collinear approximation}~\cite{RKELLIS}. This method assumes the 
neutrinos are collinear with the decaying $\tau$-lepton, and includes $\MET$ by 
projecting it to the $\tau$-jets. However, this projection is only
possible and reliable in a fraction of events with large transverse
boost and well aligned $\MET$~\cite{WU,ALETTE}. 

In \cite{MMC} the performance of a likelihood based technique is presented for
simulations of Higgs and \z0\ events in antiproton-proton collisions at
1.96~TeV, and also for data collected by the CDF experiment. For fully
hadronic decaying $tau$-leptons, unsmeared simulations show a mass
reconstruction resolution of 8\%, and more realistic simulation shows
a mass resolution of 14\%, and finally the CDF $\z0\to\tautau$ events
show a   resolution of 16\% with this technique. The smeared results
are obtained for events within $|\eta |<1$, smearing of hadron momenta
of 10\% and of $\MET$ of 5~GeV in each of the transverse components.  

At the LHC the events are much more boosted, and the experiments have
larger $\eta$ coverage. In~\cite{CMS}, CMS expects a likelihood mass
resolution of 21\% for a 130~GeV Higgs (for leptonic and semi-leptonic
$\tau$-decays), in an analysis covering $|\eta |<2.3$ for hadronic $\tau$
decays. This is approaching the width of the the \mvis\ distribution
which is reported to be 24\%. A possible explanation for the poorer
resolution in CMS could be that $\MET$ measures a smaller fraction of
the full neutrino energies for $\tau$-jets with large $\eta$ than for
transverse ones. Hence, estimates of the full neutrino energies from
$\MET$ arises from correspondingly larger scale factors.

Another mass variable for $\tau$ pairs is the effective mass,
$\meff=\sqrt{(p_{\tau^1} + p_{\tau^2} +\slashed{p}_\mathrm{T})^{2}} $, where
$p_\tau$ are the visible $\tau$-jet 4-momentum, and $\slashed{p}_\mathrm{T}$ is
the missing transverse momentum vector. In addition to the \mvis\ and
\meff\ recent result from the ATLAS collaboration~\cite{recentATLAS}
has also used a likelihood based method, whereas the recent result
from the CMS collaboration~\cite{recentCMS} simply uses \mvis\ in the
analysis.  

In the following an alternative variable for mass estimation is described and 
for simplicity comparisons have been made to distributions of \mvis\ which is 
a straightforward robust alternative in use by all experiments. 

\subsection*{Finding the kinematic edge}
\begin{figure}[htbp]
  \begin{center}
    \subfigure[Distributions for \z0 events.]{
      \includegraphics[width=0.45\textwidth]{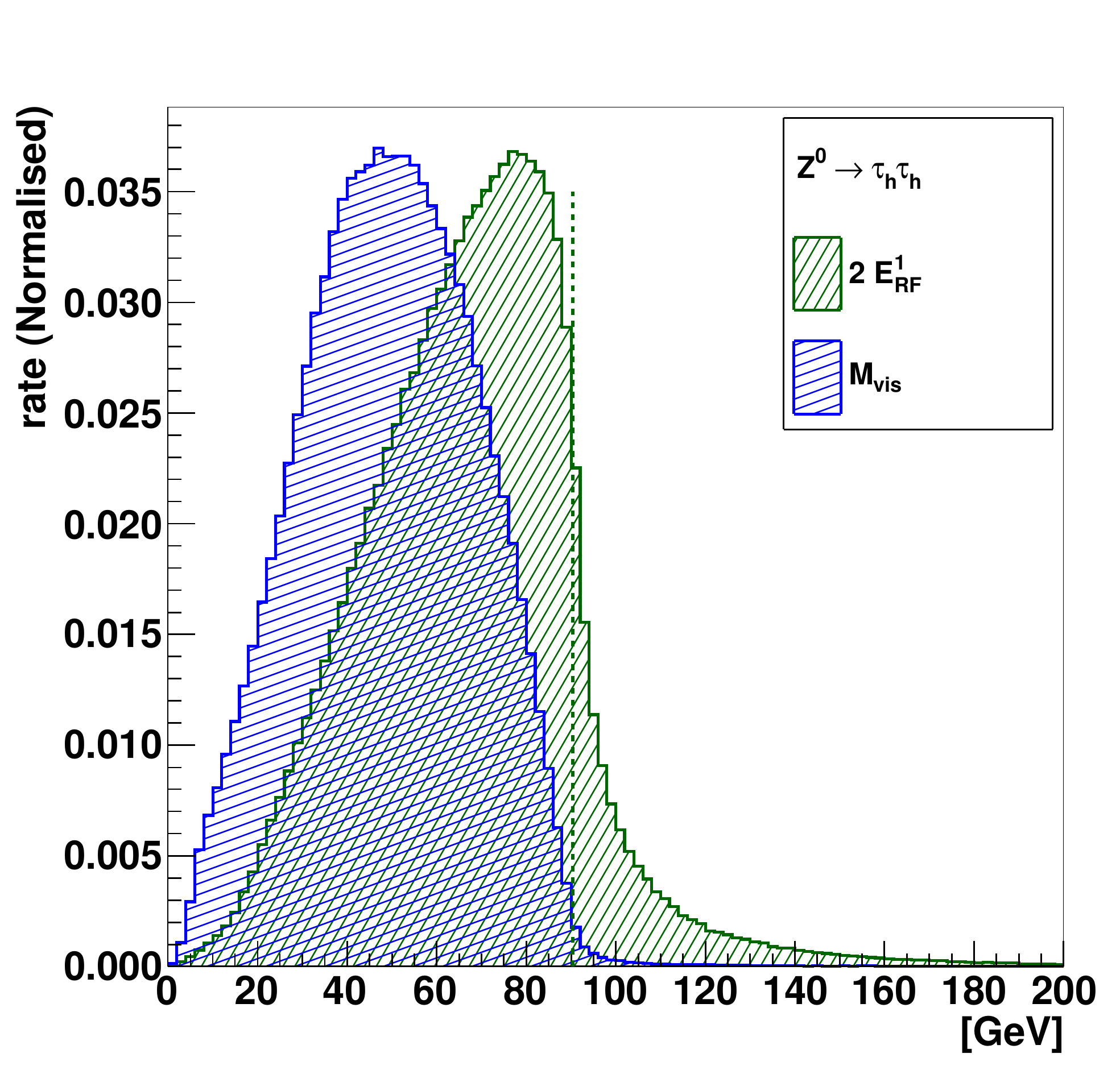}
      \label{fig:BoostMass_Z}
    }
    \hspace{0.5 cm}
    \subfigure[Distributions for $H$ events.]{
      \includegraphics[width=0.45\textwidth]{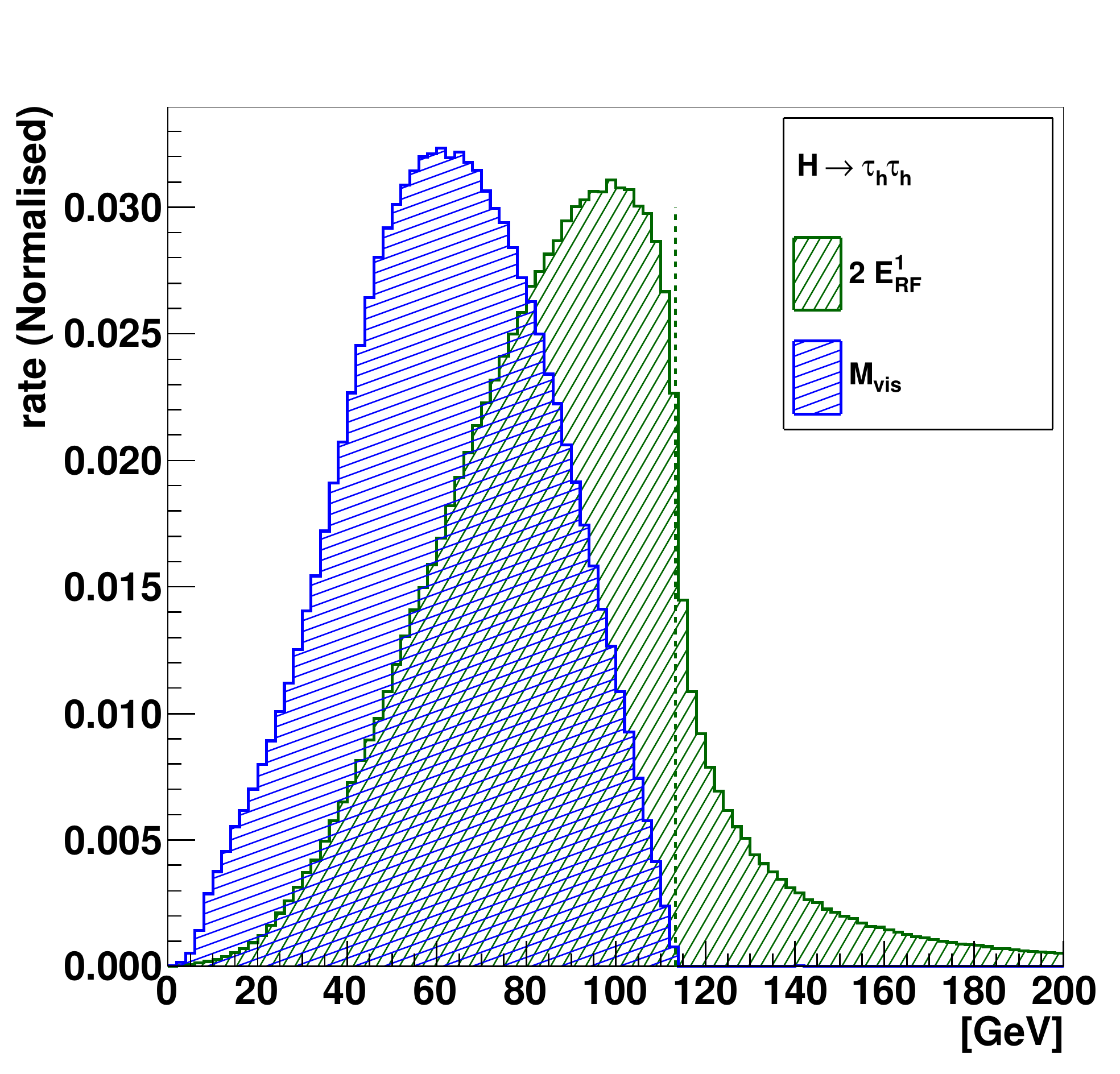}
    }
    \caption{Distributions for \eboost\ calculated 
      using the \xyz-method and 
      invariant mass of the visible $\tau$-jets, \mvis, 
      for \tautau\ pairs both decaying to single charged hadrons. 
      The dotted line marks the position of the steepest slope for 
      the \eboost\ distribution.
      \label{fig:BoostMass_H}
    }
    \label{fig:BoostMass}
  \end{center}
\end{figure}

The mass of a particle decaying to \tautau\ could be inferred by determining 
the kinematic endpoint of the distribution of twice the leading $\tau$-jet 
energy in the boson \RF. This quantity will henceforth be named as \eboost. 
As opposed to the collinear approximation this quantity can be calculated for 
all events. In figure~\ref{fig:BoostMass} distributions of 
\eboost\ are compared to \mvis\ distributions. 

This way of estimating the mass becomes particularly useful when both 
$\tau$-leptons decay into hadrons, where only one neutrino is present in each 
$\tau$ decay. 
Figure~\ref{fig:BoostMass} suggests that a first guess to the shape of \eboost\ 
could be a triangle convoluted with some resolution function, although it is 
clear that the inclusion of effects like the $\z0/H$ width and helicity effects 
modifies such an assumption. Still the kinematic edge can be found be 
determining the point of steepest slopes of the distribution. 

\begin{figure}[htbp]
  \begin{center}
    \subfigure[Class 1: Distributions for \z0 events with $\dphi>170^{\circ}$.]{
      \includegraphics[width=0.45\textwidth]{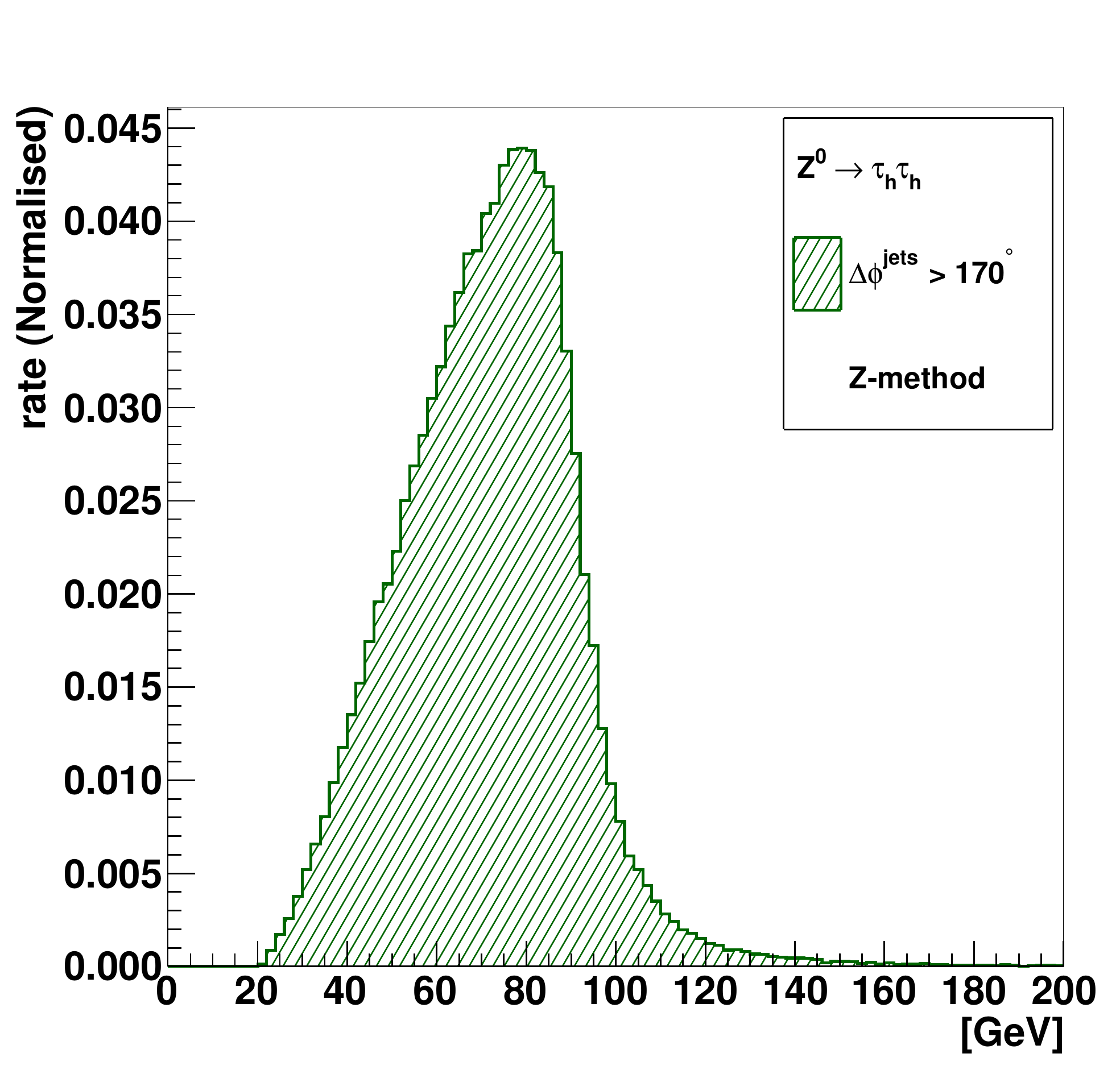}
      \label{fig:BoostMass_Z_over}
    }
    \hspace{0.5 cm}
    \subfigure[Class 2: Distributions for \z0 events with $\dphi<170^{\circ}$.]{
      \includegraphics[width=0.45\textwidth]{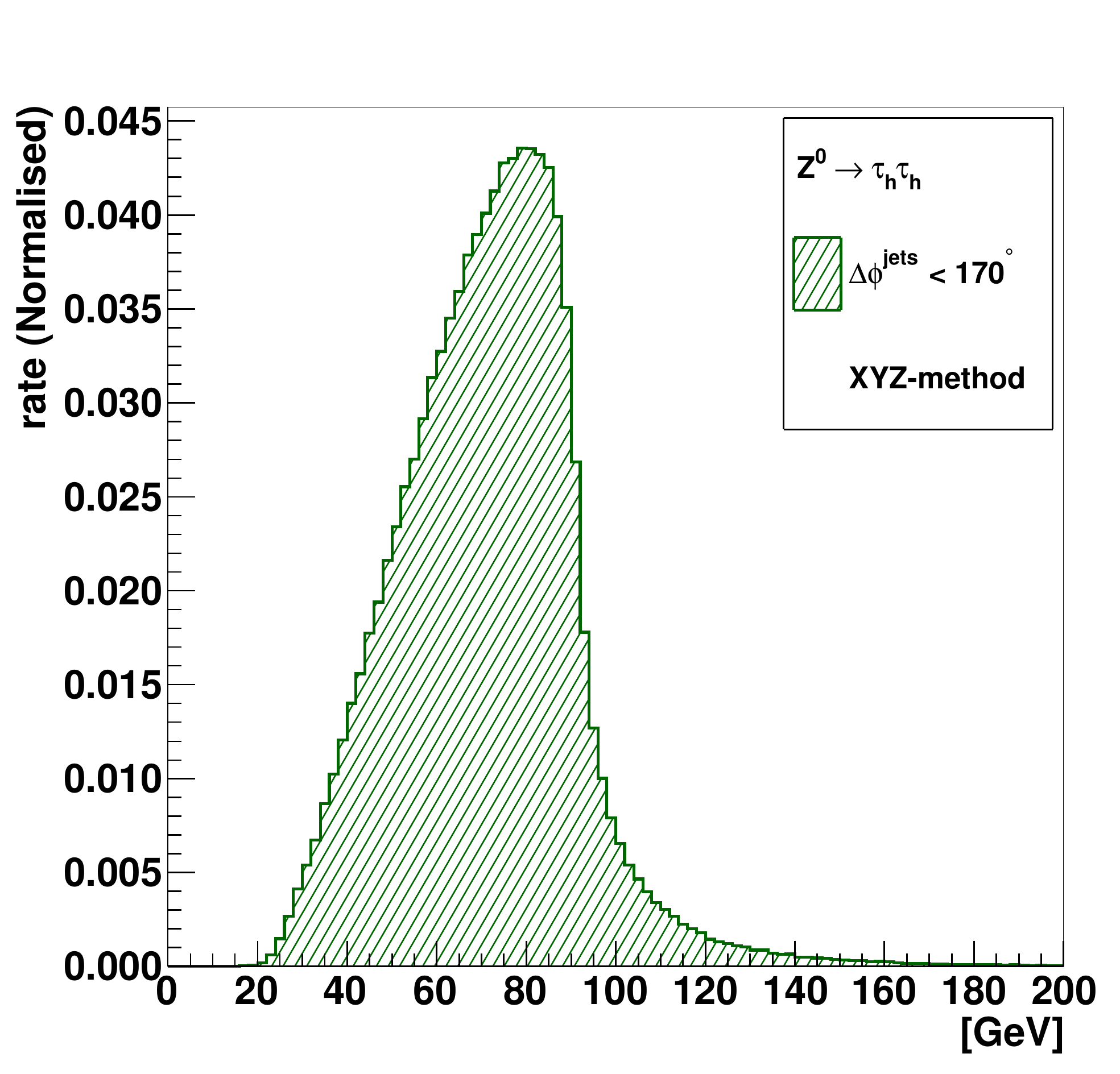}
    }
    \caption{
      Distributions for \eboost\ calculated 
      for the two event classes using the Z and \xyz-method for
      \tautau\ pairs both decaying to single charged hadrons. 
      \label{fig:BoostMass_Z_under}
    }
    \label{fig:BoostMass_cut}
  \end{center}
\end{figure}

Figure~\ref{fig:BoostMass} is the result for \emph{all} hadronic
decaying events without any selection. As mentioned earlier, the
decision on which boost reconstruction method to use 
should depend on the magnitude of the transverse
boost. Therefore, the events have been divided into two classes; one
with a large difference in azimuthal angle between the 
two jets, \dphi, compatible with no transverse boost, where the
Z-method were used, and one with the rest of the event where the
\xyz-method were applied. Here a threshold of $\dphi=170^{\circ}$ were
chosen to separate the two classes. The criterion for the choice of boosting
method would have to be subject to study in a true experimental
environment. 
Furthermore, a cut of 10~GeV on the \pt\ for both $\tau$-jets were
applied to both classes. In figure~\ref{fig:BoostMass_cut}, it is
shown that 24\% of the \z0-boson and 16\% of the Higgs events can be
reasonably subjected to the Z-method. Hence, without any use
of $\MET$ an approximation of the centre of mass system is still
available of these events. 

Since the distributions of \eboost\ are clearly asymmetric and
non-gaussian, the spread of the distributions calculated from the most 
likely value are therefore also asymmetric. For Higgs events with \dphi\ above
$170^{\circ}$, the spread from the peak position is found to be ${}^{+17}_{-25}\%$.

To evaluate the precision of retrieving the boson mass by finding the kinematic 
edge, \mboost, pseudo-experiments were performed. From the \eboost\ 
distribution shown in figure~\ref{fig:BoostMass_Z} several pseudo-distributions 
were generated with varying number of events and for each distribution the 
kinematic edge was found as the point of the steepest decent. 
In table~\ref{tab:MassError}, the found values and 
spread of \mboost\ are listed as a function of the number of events in the 
pseudo-experiment for the two classes of events, where class 1 refers to the
events in which the Z-method has been used, and class 2, refers to the
events where the \xyz-method were used. For the simple kinematic edge-finder
implemented for this analysis the \mboost\ converges to
$90.8 \pm 1.4~\GeV$ for 4000 events (for class 1 events). These values
indicates that the method could be of interest to investigate further
in experimental data.  

For a triangle function convoluted with a gaussian, numerical
calculations have shown that taking the kinematic edge to be the
point of the steepest slope overestimates the edge value. For a
gaussian width of 7--10~$\GeV$, the systematic shift is found to be  
0.5--1.0~$\GeV$. For an experimental analysis including backgrounds a
more sophisticated edge finding method, e.g. template fitting, should
be deployed. 

\begin{table}[htdp]
  \caption{Mass estimation \mboost\ as a function of number
    of events $n_\mathrm{evt}$ as obtained from pseudo-experiments for
    the two classes of \z0\ events. Class 1 consists of events with 
    $\dphi > 170^{\circ}$ where the Z-method have been applied and class 2
    consists of events that has been subject to the \xyz-method}
  \begin{center}
    \begin{tabular}{| l | c | c |}
      \hline
      & \multicolumn{2}{|c|}{Obtained values of \mboost } \\ 
      \hline 
      Number of events & Class 1 & Class 2\\
      \hline 
      250 & 88.3 $\pm$ 5.3&   88.5 $\pm$ 4.7 \\ 
      500 & 89.3 $\pm$ 3.1& 89.3 $\pm$ 2.5 \\ 
      1000 & 90.3 $\pm$ 2.3 &  89.7 $\pm$ 1.7 \\
      2000 & 90.6 $\pm$ 1.8 & 89.9 $\pm$ 1.3 \\ 
      4000 & 90.8 $\pm$ 1.4 & 90.0 $\pm$1.1 \\ 
     \hline
    \end{tabular}
  \end{center}
  \label{tab:MassError}
\end{table}

\begin{figure}[htbp]
  \begin{center}
    \includegraphics[width=0.45\textwidth]{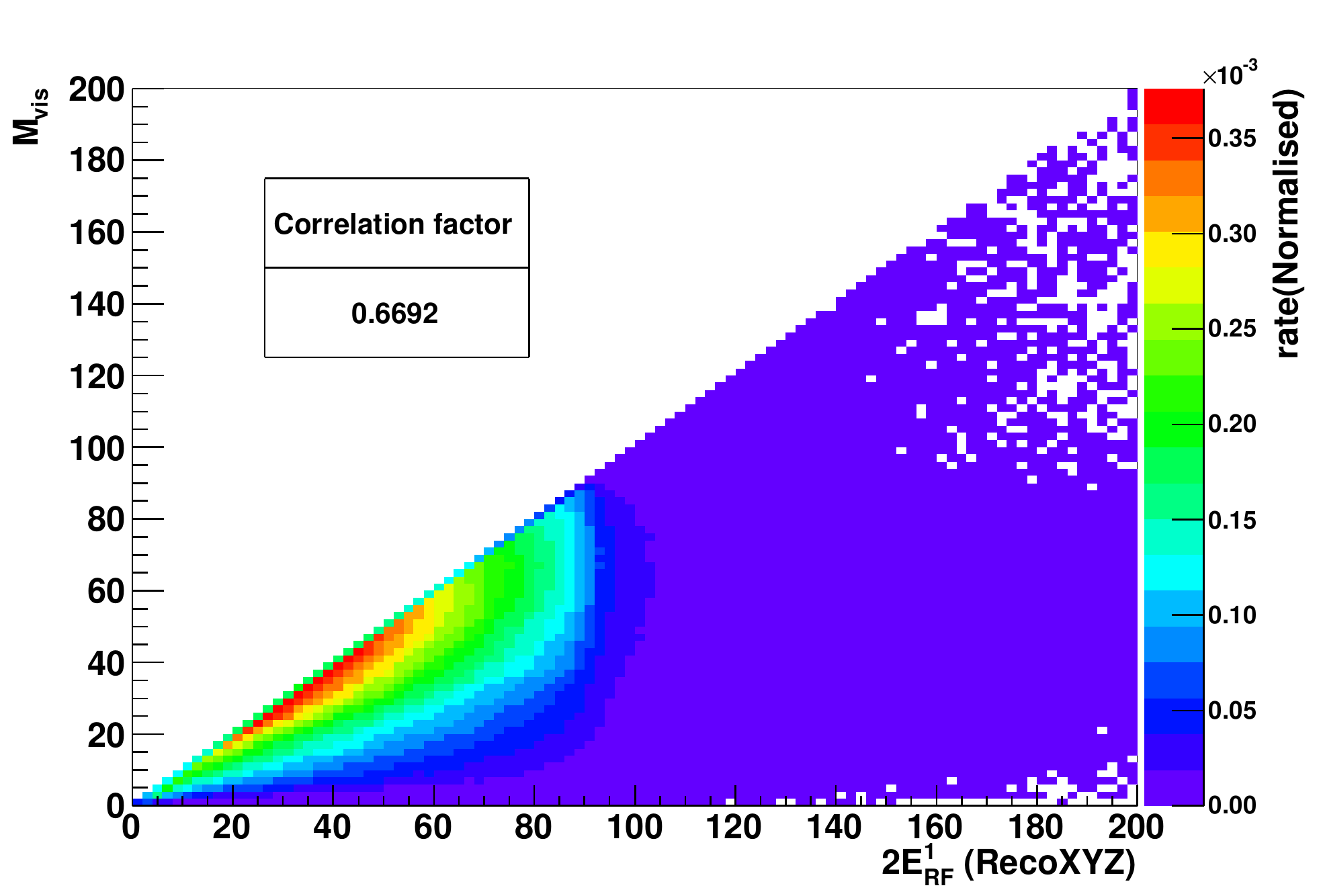}
    \caption{Correlations between \mvis\ and \eboost\ using the \xyz-method 
      in $\z0\to\tautau$ events with all decay channels of the $\tau$-leptons 
      included.}
    \label{fig:corr_mass}
  \end{center}
\end{figure}

Intuitively, constructing a mass variable that accumulates towards the real 
mass should help distinguishing different masses. One way to distinguish 
distributions is to compare the largest difference of the cumulative 
distributions, $D_L$ which is input to the Kolmogorov-Smirnov test. 
It is found that while $D_L$ between the calculated \eboost\ distributions 
(using the \xyz-method) for \z0\ and $H$ events is $0.35$, it is only $0.24$ 
when comparing the two \mvis\ distributions.
It is not clear if the tails towards high values of \eboost\ seen to be more 
important in figure~\ref{fig:BoostMass_H} than figure~\ref{fig:BoostMass_Z} is 
helpful in a search scenario, because clearly the \z0\ contribution is 
non-negligible for values well beyond the kinematic edge. At present, it cannot 
be concluded that this variable will be better than any of the other 
proposition in a real experiment. However it seems to be worthwhile to consider 
the use this variable in an experimental search for the Higgs boson.

In figure~\ref{fig:corr_mass}, the correlation between \mvis\ and \eboost\ is 
shown for \z0 events. Thus, since the correlation is significantly lower than 
1, the \eboost\ variable is adding information, and it should be studied for 
use in a search scenario. 

\section{Helicity correlations}
Several powerful variables for studying $\tau$ polarisation in \z0\ decays were 
developed for the LEP experiments~\cite{DAVIER}. These variables are defined in
the rest frame of the decaying boson. While this is close to identical to the 
laboratory frame at LEP, this is not true at the LHC, and this paper proposes 
to study these variables in a reconstructed \RF\ instead. 

For $\tau^\pm\ra\pionmode$ events, the distribution of the decay angle in the 
$\tau$-lepton rest frame, that is the angle between the direction of the 
$\tau$-lepton and its visible decay product, is determined by the $\tau$ 
helicity. 
In the rest frame of the heavy boson producing a $\tautau$ pair this 
translates to a well determined distribution in the fraction of $\tau$ energy 
carried by $\pi$. The quantity used in the following figures is the energy of 
visible particles in the decay, \Evis, normalised to the kinematic endpoint 
(half the mass of the decaying boson). Weaker energy-energy correlations also 
shows for the $\tau^\pm\ra\lmode$ and $\tau^\pm\ra\rhomode$ modes. 

\begin{figure}[htbp]
  \begin{center}
    \includegraphics[width=0.55\textwidth]{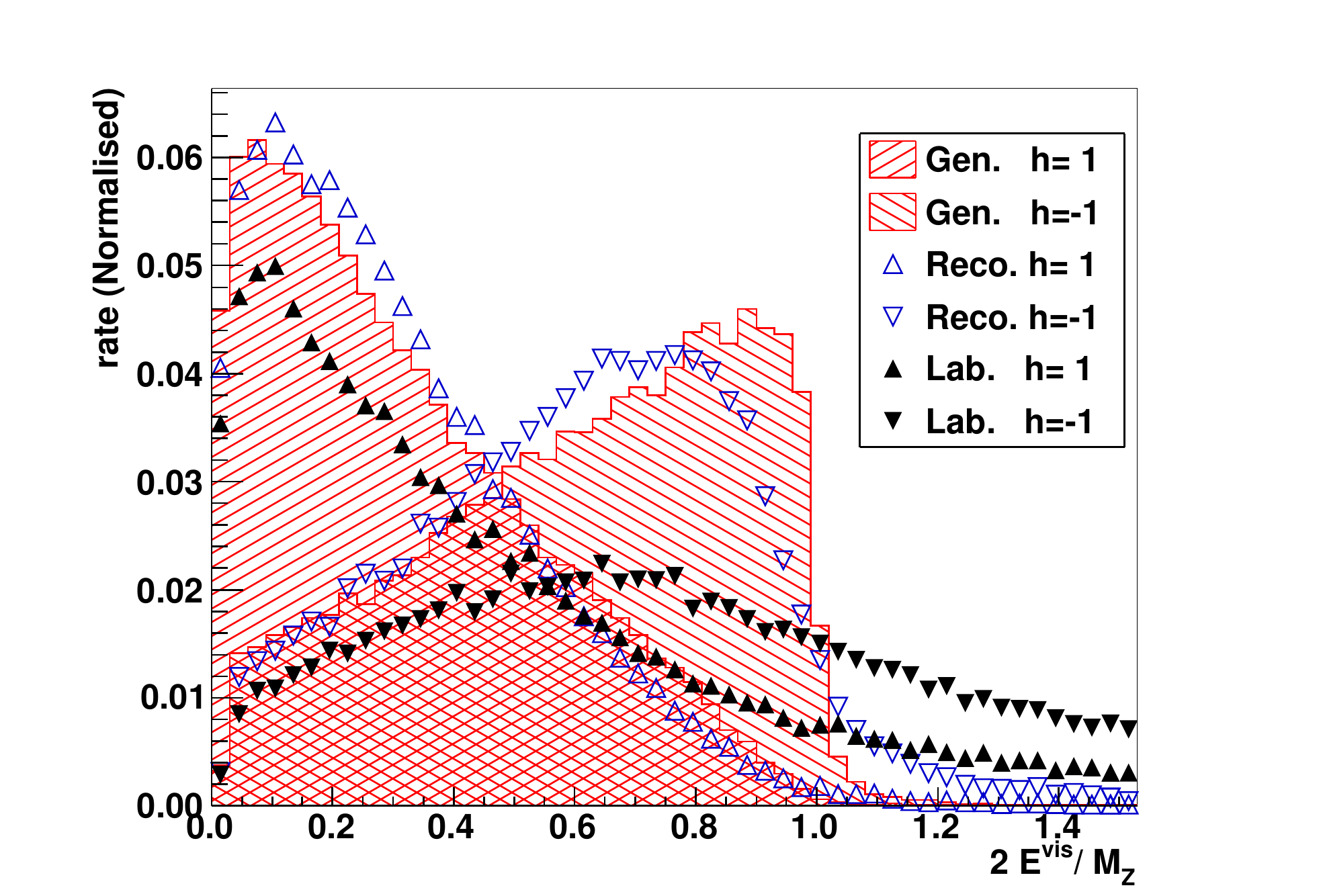}
    \caption{$2\Evis/M_{Z}$ for $\tau$-leptons with positive and negative 
      helicities decaying into $\pi^{\pm}\nu_{\tau}$ shown in the generated 
      and reconstructed \RF\ using the \xyz-method as well 
      as the laboratory frame.}
    \label{fig:pol}
  \end{center}
\end{figure}

The $\Evis$ distributions for single $\pi$ decay modes shown in 
figure~\ref{fig:pol} for the generated and reconstructed heavy boson $\RF$s 
show that the true correlations are partly recovered in the reconstructed 
\RF\ using the \xyz-method proposed in section~\ref{sec:method}. Furthermore, 
it is shown that these correlations are much weakened in the laboratory 
frame.

\subsection{Analysis of the sensitivity to spin}
\begin{figure}[htbp]
  \begin{center}
    \subfigure[Spin 0 ($H$ sample).]{
      \includegraphics[width=0.45\textwidth]{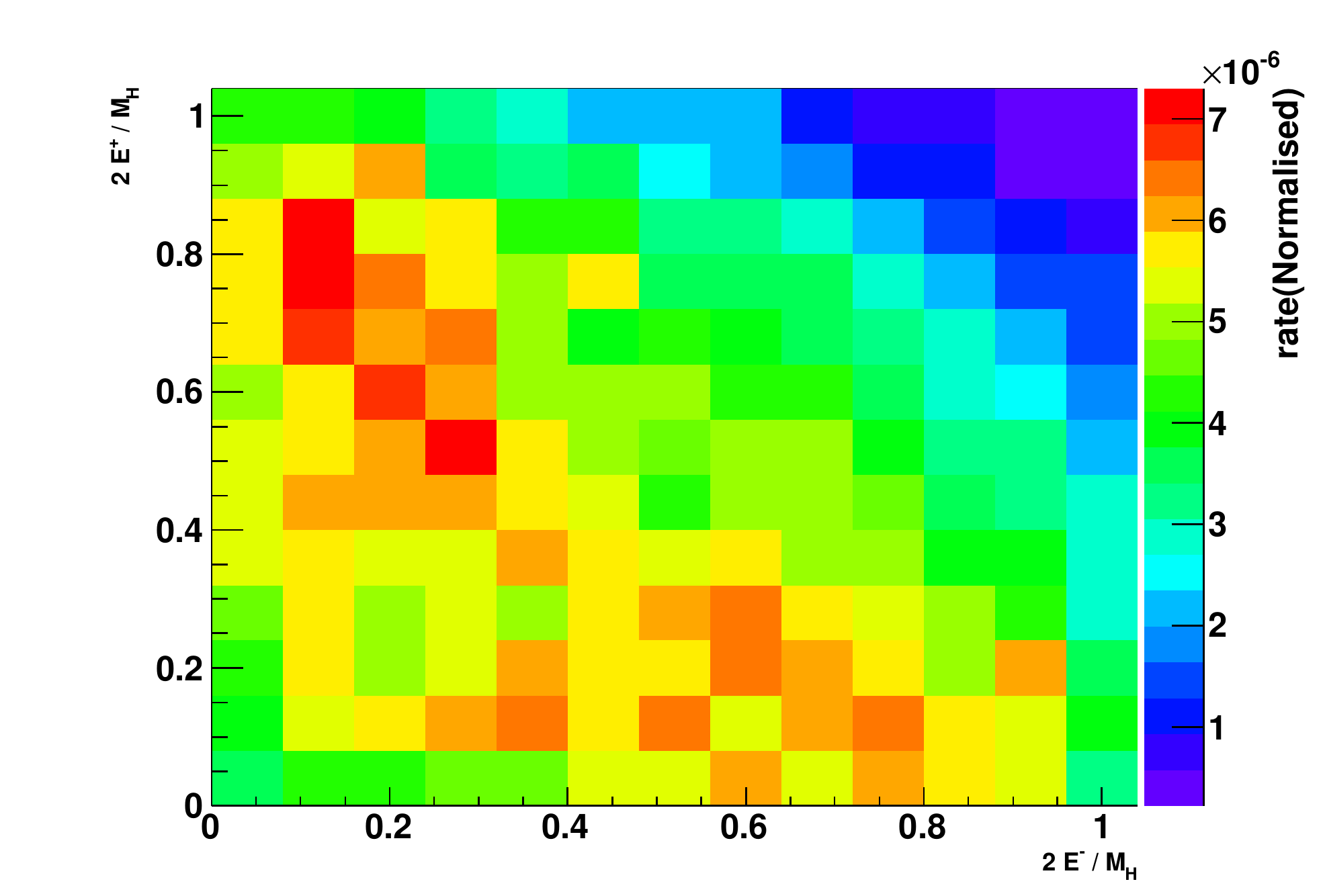}
    }
    \hspace{0.5cm}
    \subfigure[Spin 1 (\z0\ sample).]{
      \includegraphics[width=0.45\textwidth]{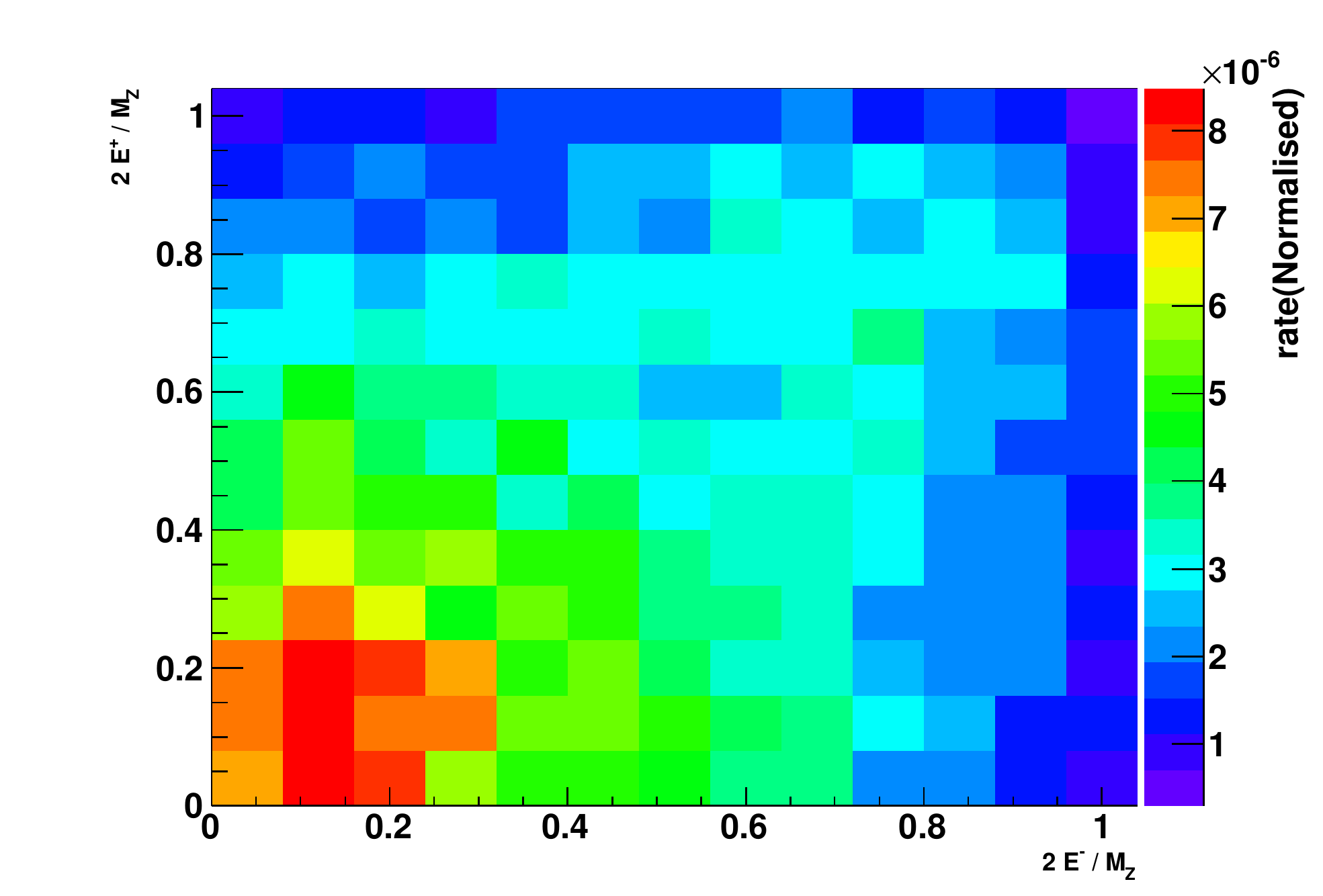}
    }
    \caption{Energy correlations in the reconstructed \RF\ using the \xyz-method 
      of a \tautau\ pair with both $\tau$-leptons decaying to \pimode. 
      To avoid effects coming from the mass differences all energies are scaled 
      with the mass of the decaying boson.}
    \label{fig:normedEs}
  \end{center}
\end{figure}

Conservation of angular momentum leads to distinctly different helicity 
configurations when comparing the final states of \z0 and $H$ decays in their 
rest frames. While the spin components along the direction of flight of the  
$\tau^{+}$ must add up to $\pm1$ for the \z0, the sum must be 0 for the $H$
decays. Thus the spin effects can be studied by looking at the correlations 
in $\Evis$ of the two $\tau$-leptons as shown in 
figure~\ref{fig:normedEs}.

To quantify the observed difference between spin 0 and spin 1 particles, the 
\z0\ and $H$ samples were split into training and test samples. Probability 
density functions for both spins, 
$P_0(E_{\tau^-},E_{\tau^+})$ and $P_1(E_{\tau^-},E_{\tau^+})$,
were constructed as the fraction of events in a bin of 
$(E_{\tau^-}, E_{\tau^+})$ in the corresponding training sample. 
From this a likelihood, $\mathcal{L}$, is created for $N$ events from both 
training samples by
\begin{equation}
  \logL = \sum_{i=0}^{N} \left( \log ( P_{1}^{i} ) - \log ( P_{0}^{i} ) \right) / N 
  \label{likelihood}
\end{equation}

\begin{figure}[htbp]
  \begin{center}
    \includegraphics[width=0.55\textwidth]{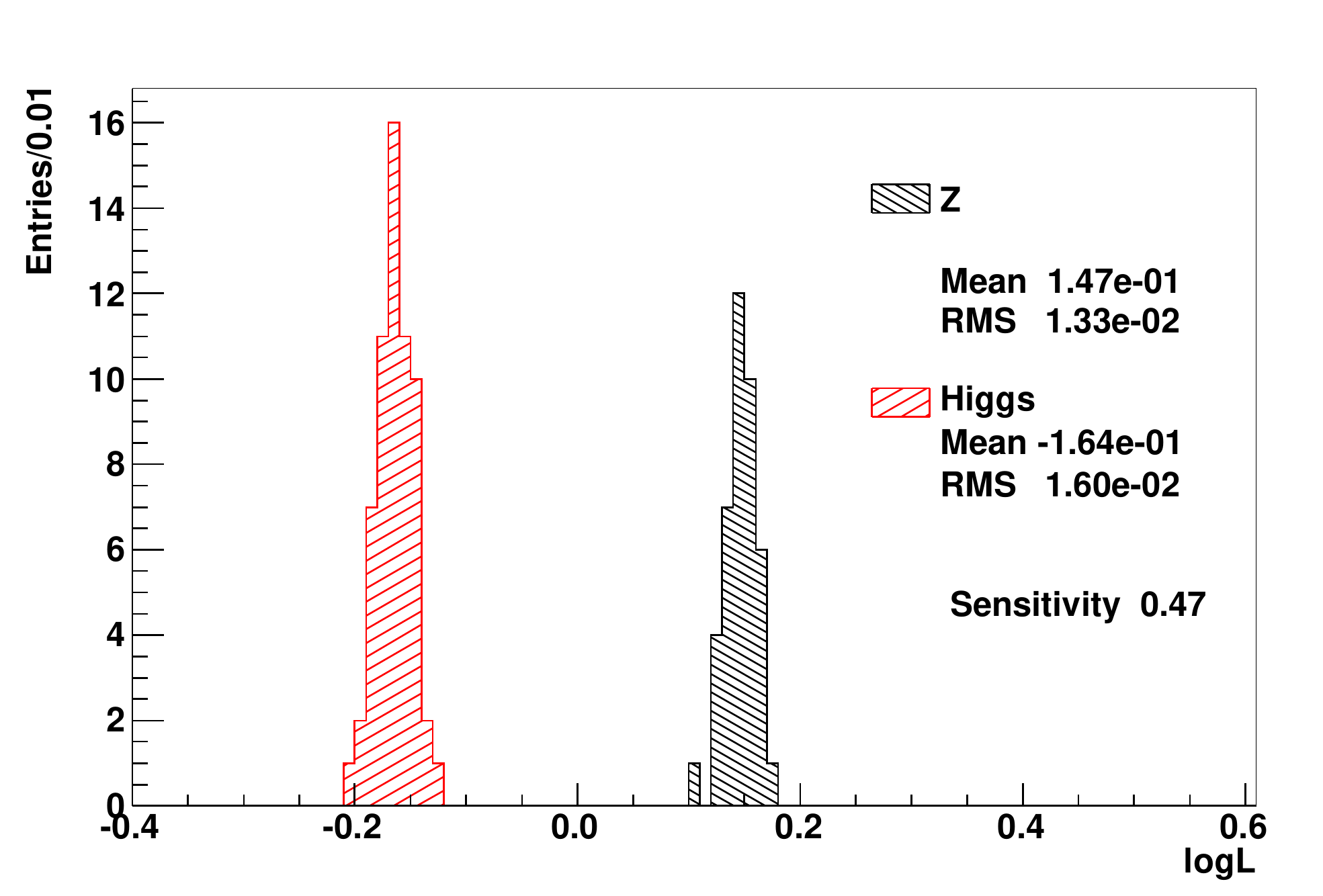}
    \caption{Log likelihood plot for energy-energy correlations in the 
      reconstructed \RF\ using the \xyz-method when both 
      $\tau$-leptons decays into \pimode.}
    \label{fig:sensitivity}
  \end{center}
\end{figure}

The test samples of $\z0$ and $H$ events are divided in sub-samples of $N$ 
events and the likelihood is calculated for each sub-sample. The distribution 
of $\logL$ when $N=1000$ and when both $\tau$-lepton decays to $\pimode$ is 
shown in figure~\ref{fig:sensitivity}. 

Inspired by \cite{DAVIER}, a \emph{sensitivity}, $S$, is defined such that the 
number of standard deviations between the means of the two distributions, 
$n_\sigma$, is given by
\begin{equation}
  n_{\sigma} = S \sqrt{N} 
\end{equation}
Table~\ref{tab:sensitivity} shows values obtained for $S$ in the reconstructed 
$\RF$s using the Z- and \xyz-method as well as in the laboratory frame for 
the different $\tau$ decay modes. For all decay channels sensitivity is 
gained when using the reconstructed \RF\ over the laboratory frame.
When just applying the Z-method, differences in the transverse momentum of the 
bosons are kept in the distributions, while the \xyz-boost approximates the 
\RF\ more correctly. The Z-method thus analyses a combination of the two 
effects, the boson $p_T$ and spin, while the sensitivities determined using the 
\xyz-method are largely only due to differences in spin.

\begin{table}[htdp]
  \caption{Sensitivity to spin of the boson for different decay modes of 
    the two $\tau$-leptons shown in the true and reconstructed $\RF$s as well as
    in the laboratory frame.}
  \begin{center}
    \begin{tabular}{| l l | c | c | c | c |}
      \hline
      \multicolumn{2}{|c|}{$\tau$ decay mode} & \multicolumn{4}{|c|}{Sensitivity } \\ 
      \hline 
      $\tau_1$ & $\tau_2$ & Reco.\RF(Z) & Reco.\RF(\xyz) & Lab. & True \RF \\ 
      \hline
      $\lmode$ & $\lmode$        & 0.13  & 0.09 & 0.06 & 0.08 \\
     $\lmode$ & $\pimode$       & 0.17  & 0.13 & 0.06 & 0.15  \\
      $\lmode$ & $\rhomode$      & 0.15  & 0.07 & 0.05 & 0.13 \\
      $\pimode$ & $\pimode $     & 0.42  & 0.47 & 0.23 & 0.51 \\
      $\pimode$ & $\rhomode$     & 0.21  & 0.15 & 0.10 & 0.18 \\
      $\rhomode$ & $\rhomode$    & 0.19  & 0.13 & 0.09 & 0.18 \\
      \hline
    \end{tabular}
  \end{center}
  \label{tab:sensitivity}
\end{table}
Finally, further sensitivity would be gained by using ratios between neutral 
and charged energy for $\rhomode$ final states. An exhaustive analysis of spin 
should probably make use of these ratios. These ratios remain almost unchanged 
in any boosted frame. The purpose of this note is to discuss the advantages of 
finding an approximate rest frame, so the sensitivities when using these 
variable are not reported here.

\section{Conclusions and discussion}
By minimising the acollinearity, two methods to reconstruct the rest frame of 
boosted heavy particles decaying to two $\tau$-leptons have been proposed, 
and it has been demonstrated that they both find the rest frame reasonably 
well. Whereas the \xyz-method requires an estimate of the transverse direction 
of the heavy particle, the Z-method can be used for analyses where one does 
not have a good estimate of this direction.

A new technique for mass estimation has been proposed using the end-point of 
the leading $\tau$-jet energy in the reconstructed rest frame. This technique 
does not introduce assumptions on transfer functions or production mechanisms 
of the heavy particle as more elaborate reconstruction techniques, and it can 
be applied to \emph{all} events within the geometrical 
acceptance of the experiments. The method is complementary to existing
methods in the sense that it can be applied without using information
on the measured $\MET$, and since it is not directly correlated with
the visible mass it adds information, and thus the two can be combined
to gain further sensitivity.

Furthermore, a scheme for an event-wise spin analysis of the \tautau\ system 
has been outlined, and it has been shown that enhanced sensitivity to spin is 
gained when transforming the 4-momenta as proposed in this paper. Both boost 
methods presented enhances sensitivity to spin when compared to an analysis 
in the laboratory frame. Which method to use depends on whether or not 
differences in production mechanism of the boson should be taken into account.       

In summary, a tool has been developed to approximate the rest frame of \tautau\ 
systems, and two applications thereof have been illustrated. These applications 
could prove to be helpful in the analyses of \z0 decays or in searches for new 
particles like $H$, $Z^\prime$ or SUSY particles. Additionally, it is possible 
that a transformation to the rest frame could be beneficial to other variables 
used in the study of \tautau\ systems.

\bibliographystyle{JHEP}
\bibliography{boostnote}
\end{document}